\documentstyle[draft,aps,epsfig]{revtex}
\begin{document}
\draft
\tightenlines
\title{Isospin-Breaking Vacuum-to-$\pi^0,\eta$ Pseudoscalar Matrix \\
Elements at Next-to-Leading Order in the Chiral Expansion}
\author{Kim Maltman} 
\address{Department of Mathematics and Statistics, York University \\
          4700 Keele St., North York, Ont. CANADA M3J 1P3}
\address{Special Research Center for the Subatomic Structure of Matter \\
          University of Adelaide, Australia 5005}
\author{Carl E. Wolfe}
\address{Department of Physics and Astronomy, York University \\
          4700 Keele St., North York, Ont. CANADA M3J 1P3}
\date{\today}
\maketitle
\begin{abstract}
We employ Chiral Perturbation Theory (ChPT) to evaluate the complete set of 
pseudoscalar matrix elements, $<0|P_f|\pi^0 ,\eta>$, with $P_f$ 
any of the flavor-diagonal pseudoscalar currents ($f=u,d,s$), to 
${\cal O}(m_d-m_u)$, and to next-to-leading order in the chiral expansion.
These matrix elements represent the basic input to a QCD sum rule
analysis of isospin breaking in the $\pi NN$ couplings using the 
three-point-function method.  
We discuss also how one could use the results to construct a one parameter
family of interpolating fields for the $\pi^0$, all of whose members have
zero vacuum-to-$\eta$ matrix element, and explain how this could in principle
be used to provide non-trivial tests of the reliability
of the assumptions underlying the use of the three-point-function
method.  It is shown that the isospin-breaking mixing parameters required
for this construction receive significant corrections beyond leading
order in the chiral expansion.
\end{abstract}

\section{Introduction}
QCD sum rules\cite{svz} treatments of the isospin-conserving $\pi NN$ coupling,
both using the two-point-function and three-point-function
methods\onlinecite{ref1,ref2,ref3,ref4,ref5,ref6,ref7}, have proven
quite successful.  As a consequence, it is natural to consider
extending these treatments to an investigation of the strong 
interaction ($m_d-m_u\not= 0$) contributions to the isospin breaking of
these same couplings.  The resulting differences among
$g_{pp\pi^0}$, $g_{nn\pi^0}$ and $g_{np\pi^+}$ would provide
contributions to isospin-breaking observables in few-body systems, and an 
evaluation based on methods closely related to QCD is, therefore,
highly desirable, particularly in light of the recent revival of interest
in such phenomena (especially charge symmetry breaking 
(CSB))\onlinecite{mnsrev,ght,pw93,gtw93,mg94,chm,hhkm,krmps,krm12,kent,chptcsb,ghp,vkgf,ijl,cm95,mow,hm95}.

A first attempt in this direction was made in Ref.\onlinecite{hm95} (MH),
where a sum rule analysis using the three-point-function method was
performed.  In this approach, one considers the three-point
functions
\begin{equation}
A_{N^\prime N\pi}(p_1,p_2,q)= \int\, d^4x_1\, d^4x_2\, \exp [{\rm i}(
p_1.x_1-p_2.x_2)]\, <0|T\left( \eta_{N^\prime}(x_1)
P_\pi (0) \overline{\eta}_N(x_2)\right)|0>\label{3pt}
\end{equation}
where $\eta_{N,N^\prime}$ are (Ioffe) interpolating fields for the
nucleons $N^\prime$, $N$ and $P_\pi$ a pseudoscalar interpolating
field for the pion of interest (which carries momentum $q=p_1-p_2$).
The desired coupling, $g_{N^\prime N\pi}$, then appears as a coefficient
in the contribution to the phenomenological (``hadronic'') representation
of $A_{N^\prime N\pi}$ having simultaneous
poles at $p_1^2=p_2^2=M_N^2$ and $q^2=m^2_\pi$.  To evaluate these
couplings in terms of vacuum condensates, one then simultaneously evaluates
$A_{N^\prime N\pi}$ using the operator product expansion (OPE) (the
validity of which requires $q^2< -1$ GeV$^2$), and relates the
hadronic and OPE representations of the correlator via an 
appropriately Borel transformed dispersion relation\onlinecite{svz}.  
The contributions due
to the pion coupling are isolated by keeping only those terms
in the OPE proportional to $\not\!{q}\gamma_5$ (as in the 
phenomenological representation) which have a $1/q^2$ pole in 
the limit that the quark masses are neglected\onlinecite{ref1,ref2,ref3,hm95}.

It is necessary to exercise some care in the choice of the interpolating
field for the neutral pion in Eq.~(\ref{3pt}).  Indeed, if one were to 
make the most
obvious choice, {\it i.e.}, the $I=1$ combination
\begin{equation}
P_{\pi_3}=P_{I=1}=P_u-P_d
\end{equation}
where $P_f=\bar{f} i\gamma_5 f$ is the pseudoscalar current for quark
flavor, $f$, one would find that because, to ${\cal O}(m_d-m_u)$, 
$P_{\pi_3}$ also couples to the $\eta$, the analysis for the
extraction of the isospin-breaking $\pi NN$ couplings was
contaminated by $\eta$ contributions.  This contamination is
of the same order in isospin breaking as the $\pi NN$ isospin-breaking 
couplings of interest, and hence not negligible\cite{hm95}.
This problem was overcome in MH by the alternate choice
\begin{equation}
P^{MH}_{\pi^0}=(P_u-P_d)\, +\, \epsilon^{MH}P_8\ ,
\label{mhpi0}
\end{equation}
where
\begin{equation}
P_8={\frac{1}{\sqrt 3}}\left( P_u+P_d-2P_s\right)
\end{equation}
and $\epsilon^{MH}$ is to be chosen such that $<0|P^{MH}_{\pi^0}|\eta >=0$.
To leading order in the chiral expansion, one has, from the results of
Gasser and Leutwyler\onlinecite{gl85},
\begin{equation}
\epsilon^{MH}=\theta_0={\frac{\sqrt 3}{4}}\left(
{\frac{m_d-m_u}{m_s-\hat{m}}}\right) \ ,
\end{equation}
where $\theta_0$ is the leading order $\pi^0-\eta$ mixing angle.
Since, however, it is known that next-to-leading order contributions
to, {\it e.g.}, $f_\eta$ are large ($\sim 30\%$), it is likely that
similar next-to-leading order corrections to $\epsilon^{MH}$ will
also be important.

In this paper we will compute the pseudoscalar current matrix elements,
$<0|P_f|\pi^0 ,\eta>$, required to evaluate these
corrections, to ${\cal O}(m_d-m_u)$ and to next-to-leading
(1-loop) order in the chiral expansion.
We will, in addition, show that it is possible to use
these matrix elements to construct, not just the field $P_{\pi^0}^{MH}$
above, but a continuous family of similar interpolating fields, all
with zero vacuum-to-$\eta$ matrix element.  These fields can then,
in principle, be used to study the reliability of the assumptions
which go into the use of the three-point-function treatment of the
isospin-breaking $\pi NN$ couplings, a topic we will return to in
a later publication.

Let us consider, now, the problem of constructing ``pure'' $\pi^0$
and $\eta$ interpolating fields from the flavor diagonal pseudoscalar
currents.  If $\hat{P}_0$ is any $I=0$ pseudoscalar
field combination having non-zero $\eta$ matrix element, then forming
\begin{equation}
P_{\pi^0}=(P_u-P_d)+\hat{\epsilon}_\pi \hat{P}_0\label{interpolpi}
\end{equation}
and requiring $<0|P_{\pi^0}|\eta >=0$, {\it i.e.},
\begin{equation}
\hat{\epsilon}_\pi = -{\frac{<0|P_u-P_d|\eta >}{<0|\hat{P}_0|\eta>}},
\label{epspi}
\end{equation}
produces an interpolating field, $P_{\pi^0}$, which couples to the
physical $\pi^0$, but not at all to the $\eta$.  A ``pure'' $\eta$
interpolating field is constructed analogously:
\begin{equation}
P_\eta =\hat{P}_0+\hat{\epsilon}_\eta (P_u-P_d)\label{interpoleta}
\end{equation}
with
\begin{equation}
\hat{\epsilon}_\eta = -{\frac{<0|\hat{P}_0|\pi^0>}{<0|P_u-P_d|\pi^0 >}}\ .
\label{epseta}
\end{equation}
The basic ingredients in the construction of $P_{\pi^0}$, $P_\eta$
are, therefore, the vacuum-to-$\pi^0,\eta$ matrix elements of the $I=1$
($P_{I=1}$) and $I=0$ ($P_u+P_d$ and $P_s$) pseudoscalar currents,
where $\pi^0$, $\eta$ refer to the physical (isospin-mixed) pseudoscalar
mesons.  It is these matrix elements that we will evaluate to 
${\cal O}(m_d-m_u)$, and to 1-loop order in Chiral Perturbation Theory (ChPT)
(next-to-leading order in the chiral expansion) below.

\section{Evaluation of the vacuum-to-$\pi^0,\eta$ matrix elements of the
pseudoscalar currents}

Having explained above why an evaluation of the isospin-breaking
pseudoscalar current vacuum-to-$\pi^0,\eta$ matrix elements is of interest, 
we now turn to the explicit evaluation of these quantities, for which
we employ the methods of ChPT\onlinecite{gl85}.  In
order to obtain both the leading and next-to-leading contributions
in the chiral expansion of these quantities, it is necessary to include 
those pieces of the effective Lagrangian up to and including fourth
order in the chiral counting\onlinecite{weinberg79}, in the presence
of scalar and pseudoscalar external sources, $s$ and $p$.  The
relevant terms, as worked out by Gasser and Leutwyler\onlinecite{gl85}, are 
\begin{equation}
{\cal L}_{eff}
 = {\cal L}^{(2)}\ +\ {\cal L}^{(4)}
\end{equation}
where
\begin{equation}
{\cal L}^{(2)} = {\frac{1}{4}}f^2 \text{Tr}\left( \partial_\mu U \partial^\mu
        U^\dagger \right)
        +{\frac{1}{4}}f^2 \text{Tr}[\chi^\dagger U +\chi U^\dagger ]
\label{leff2}
\end{equation}
and
\begin{eqnarray}
{\cal L}^{(4)} &=&
        L_1 \bigl[ \text{Tr}(\partial_\mu U \partial^\mu U^\dagger )
        \bigr]^2 + L_2 \text{Tr}(\partial_\mu U \partial_\nu U^\dagger )
        \text{Tr}(\partial^\mu U \partial^\nu U^\dagger )
        + L_3 \text{Tr}(\partial_\mu U^\dagger \partial^\mu U
        \partial_\nu U^\dagger \partial^\nu U)  \nonumber \\  
     && \mbox{} + L_4 \text{Tr}(\partial_\mu U \partial^\mu U^\dagger )
        \text{Tr}(\chi^\dagger U + \chi U^\dagger ) 
        + L_5 \text{Tr}\bigl[ \partial_\mu U^\dagger \partial^\mu U 
        (\chi^\dagger U + U^\dagger \chi )\bigr] \nonumber \\
     && \mbox{} + L_6 \bigl[ \text{Tr}(\chi^\dagger U +\chi U^\dagger )\bigr]^2
        + L_7 \bigl[ \text{Tr}(\chi^\dagger U -\chi U^\dagger ) \bigr]^2 
        +L_8 \text{Tr}(\chi^\dagger U\chi^\dagger U + 
          \chi U^\dagger \chi U^\dagger )
         \nonumber \\
     && \mbox{}	 +H_2\text{Tr}(\chi^\dagger \chi)
	 \label{leff4} 
\end{eqnarray}            
where $U=\exp ({\rm i}\vec{\lambda}.\vec{\pi}/f)$, with $\{ \pi^a\}$
the octet of pseudoscalar Goldstone boson fields and $\{ \lambda^a\}$
the usual Gell-Mann matrices, $f$ is a (second order) low-energy constant
(LEC) which turns out to be equal to 
the pion decay constant in the chiral limit,
\begin{equation}
\chi = 2B_0 (s+{\rm i}\, p)
\end{equation}
with $B_0$ another second order LEC, related to the quark condensate
in the chiral limit, and $\{ L_k\}$, $H_2$ are fourth order LEC's.
Terms in the general fourth order piece of the effective Lagrangian,
${\cal L}_{eff}$,
which vanish in the absence of external vector and axial vector sources
have been dropped in writing Eqs.~(\ref{leff2}), (\ref{leff4}).
The second and fourth order LEC's appearing in these equations are 
not determined by the symmetry arguments which go into constructing 
${\cal L}_{eff}$ and must be determined phenomenologically.  (See,
for example, Ref.~\onlinecite{bkm95} for a recent up-to-date
tabulation).  For our purposes, the external scalar source is to be
set equal to the quark mass matrix, $M$, in order to correctly
incorporate the effects of explicit chiral symmetry breaking, and
the external pseudoscalar source is to be retained to first order,
in order to generate the
correct representation of the underlying QCD pseudoscalar currents 
in the low-energy effective theory.

From the form of ${\cal L}_{eff}$ given above, it is straightforward to obtain
the following low-energy representations of the pseudoscalar currents
in terms of the Goldston boson fields $\{ \pi^a\}$ occuring in $U$:
\begin{eqnarray}
P_u&=& B_0 f\Biggl[ \left(\pi_3 +{\frac{\pi_8}{\sqrt 3}}\right) 
-{\frac{1}{6f^2}}\Biggl( \pi_3^3 +\sqrt{3}\pi_3^2\pi_8 +\pi_3\pi_8^2
+ {\frac{\pi_8^3}{3\sqrt{3}}}+2\pi^+\pi^-\pi_3\nonumber \\
&&\quad +{\frac{6}{\sqrt{3}}}
\pi^+\pi^-\pi_8+4K^+K^-\pi_3
+4K^0{\bar K}^0\pi_3 +2\sqrt{2} \pi^+K^-K^0
+2\sqrt{2}\pi^-K^+{\bar K}^0\Biggr)\Biggr]\nonumber \\
&&\quad +{\frac{32B_0^2}{f}}\Biggl[ (m_s+2\hat{m})
\left(\pi_3+{\frac{\pi_8}{\sqrt{3}}}\right) L_6
+\left( (m_u-m_d)\pi_3 +{\frac{2}{\sqrt{3}}}(\hat{m}-m_s)\pi_8\right) L_7
\nonumber \\
&&\quad +m_u\left( \pi_3+{\frac {\pi_8}{\sqrt{3}}}\right) L_8 \Biggr]\ 
+\ \cdots
\nonumber \\
P_d&=& B_0 f\Biggl[ \left( -\pi_3 +{\frac{\pi_8}{\sqrt 3}}\right) 
-{\frac{1}{6f^2}}\Biggl( -\pi_3^3 +\sqrt{3}\pi_3^2\pi_8 -\pi_3\pi_8^2
+ {\frac{\pi_8^3}{3\sqrt{3}}}-2\pi^+\pi^-\pi_3\nonumber \\
&&\quad +{\frac{6}{\sqrt{3}}}
\pi^+\pi^-\pi_8
-4K^+K^-\pi_3
-4K^0{\bar K}^0\pi_3 +2\sqrt{2} \pi^+K^-K^0
+2\sqrt{2}\pi^-K^+{\bar K}^0\Biggr)\Biggr]\nonumber \\
&&\quad +{\frac{32B_0^2}{f}}\Biggl[ (m_s+2\hat{m})
\left( -\pi_3+{\frac{\pi_8}{\sqrt{3}}}\right) L_6
+\left( (m_u-m_d)\pi_3 +{\frac{2}{\sqrt{3}}}(\hat{m}-m_s)\pi_8\right) L_7
\nonumber\\ &&\quad
+m_d\left( -\pi_3 +{\frac{\pi_8}{\sqrt{3}}}\right) L_8 \Biggr]\ +\ \cdots
\nonumber \\
P_s&=& B_0 f\Biggl[ -{\frac{2}{\sqrt{3}}}\pi_8 
-{\frac{1}{6f^2}}\Biggl( -8 {\frac{\pi_8^3}{3\sqrt{3}}}
+2K^+K^-\pi_3 - 2K^0{\bar K}^0\pi_3 -2K^0{\bar K}^0\pi_3
-{\frac{6}{\sqrt{3}}}K^+K^-\pi_8
\nonumber \\
&&\quad 
-{\frac{6}{\sqrt{3}}}K^0{\bar K}^0\pi_8
+2\sqrt{2} \pi^+K^-K^0
+2\sqrt{2}\pi^-K^+{\bar K}^0\Biggr)\Biggr]
+{\frac{32B_0^2}{f}}\Biggl[ -{\frac{2}{\sqrt{3}}}(m_s+2\hat{m})
\pi_8 L_6
\nonumber \\
&&\quad 
+
\left( (m_u-m_d)\pi_3 +{\frac{2}{\sqrt{3}}}(\hat{m}-m_s)\pi_8\right) L_7
-{\frac{2}{\sqrt{3}}}m_s\pi_8 L_8 \Biggr]\ +\ \cdots
\label{currents}
\end{eqnarray}
where $\pi_3$ and $\pi_8$ are the
pure $I=1,0$ unrenormalized octet fields appearing in the matrix
variable $U$, $\hat{m}=(m_u+m_d)/2$, and we have 
written down explicitly only those terms which will be
required for our calculations below.

Using these representations, it is then straightforward to compute the
matrix elements in question.  The set of graphs which contribute is
shown in Fig.~1.  For practical purposes it is convenient to recast
the low-energy representation of the currents in terms of the renormalized, 
physical $\pi^0$, $\eta$ fields, rather than 
unrenormalized octet fields $\pi_3$ and $\pi_8$.
In order to correctly incorporate all effects at next-to-leading order
in the chiral expansion, it is necessary, for those terms linear in
$\pi_3$, $\pi_8$ and not proportional to the fourth order LEC's $L_{6,7,8}$ 
in the representations of the currents, to use a
transformation between the $\pi_3$, $\pi_8$ and $\pi^0$, $\eta$ bases
which is valid to 1-loop order.  The relevant expressions 
are\onlinecite{krmps}:
\begin{eqnarray}
\pi^0&=&Z_3^{-1/2}
\left(\pi_3+{\hat\theta}_1\pi_8\right)\nonumber \\
\eta &=&Z_8^{-1/2}
\left(-{\hat\theta}_2\pi_3+\pi_8\right)
\label{sixteen}
\end{eqnarray}
where $Z_3$ and $Z_8$ are the $\pi_3$ and $\pi_8$ wavefunction
renormalization constants to 1-loop order in the isospin symmetry
limit and
\begin{eqnarray}
{\hat\theta}_1=&&\theta_0\Biggl[ 1-{7\over 3}\mu_\pi +{4\over 3}\mu_K +\mu_\eta
+\left( {3m_\eta^2+m_\pi^2\over 64\pi^2f^2}\right) 
\left( 1+{m_\pi^2\over {\bar m}_K^2-m_\pi^2}
\log (m_\pi^2/{\bar m}_K^2)\right) \nonumber \\
&&\mbox{}+\left({m_\eta^2-m_\pi^2\over 64\pi^2f^2}\right) 
\left( 1+\log ({\bar m}_K^2/\mu^2)\right)
-{32({\bar m}_K^2-m_\pi^2)\over f^2}(3L_7^r+L_8^r)\Biggr]
\nonumber \\
{\hat\theta}_2=&&\theta_0\Biggl[ 1-{11\over 3}\mu_\pi +
{8\over 3}\mu_K +\mu_\eta
+\left({3m_\eta^2+m_\pi^2\over 64\pi^2f^2}\right) 
\left( 1+{m_\pi^2\over {\bar m}_K^2-m_\pi^2}
\log (m_\pi^2/{\bar m}_K^2)\right)\nonumber \\
&&\mbox{}-\left({m_\eta^2-m_\pi^2\over 64\pi^2f^2}\right)
\left( 1+\log ({\bar m}_K^2/\mu^2)\right)
-{32({\bar m}_K^2-m_\pi^2)\over F^2}(3L_7^r+L_8^r)\Biggr]
\ .\label{seventeen}
\end{eqnarray}
In Eqs.~(\ref{seventeen}), 
$\mu_P\equiv [m_P^2\log (m_P^2/\mu^2)]/32\pi^2f^2$, with $\mu$
the (dimensional regularization) renormalization scale, $L_k^r(\mu )$
are the renormalized version of the fourth order LEC's, and 
${\bar m}_K^2$ is the average of the neutral and charged kaon squared
masses.
Standard manipulations then
lead to the following expressions for the vacuum-to-$\pi^0,\eta$
matrix elements, valid to ${\cal O}(m_d-m_u)$, and to fourth order
in the chiral expansion:
\begin{eqnarray}
&&<0|P_u-P_d|\pi^0 >= \frac{m^2_\pi f_\pi}{\hat{m}} \label{mI1pi}\\
&&<0|P_u+P_d|\pi^0 >= \frac{m^2_\pi f_\pi\theta_0}{\sqrt{3}\hat{m}}
\Biggl[ 1+\left( {\frac{-7\mu_\pi +4\mu_K +3\mu_\eta}{f^2}}\right)
+\left({\frac{3m_\eta^2 +m_\pi^2}{64\pi^2f^2}}\right)
\Biggl( 1+\nonumber \\
&&\qquad\qquad \left[ {\frac{m_\pi^2}{m_K^2-m_\pi^2}}\right]\log\left(
{\frac{m_\pi^2}{m_K^2}}\right)\Biggr)
+{\frac{(\bar{m}_K^2-m_\pi^2)}{16\pi^2f^2}}\left(
1+\log (m_K^2/\mu^2 )\right)\nonumber \\
&&\qquad\qquad 
-{\frac{288B_0}{f^2}}(m_s-\hat{m})L_7^r -{\frac{96B_0}{f^2}}
(m_s-\hat{m})L_8^r \Biggr] \label{mlI0pi}\\
&&<0|P_s|\pi^0 >= -{ \frac{m^2_\pi f_\pi\theta_0}{\sqrt{3}\hat{m}}}
\Biggl[ 1+\left( {\frac{-2\mu_\pi +2\mu_K}{f^2}}\right)
-{\frac{B_0(m_s-\hat{m})}{16
\pi^2f^2}}\left( 1+\log (m_K^2/\mu^2)\right)\nonumber \\
&&\qquad\qquad +{\frac{B_0(m_s+\hat{m})}{16\pi^2f^2}}\left(
1+ {\frac{m^2\pi}{(m_K^2-m_\pi^2)}}\log \left({\frac{m_\pi^2}{m_K^2}}\right)
\right)\Biggr] \label{mspi}\\
&&<0|P_u-P_d|\eta >= -{ \frac{m^2_\pi f_\pi\theta_0}{\hat{m}}}
\Biggl[ 1+\left( {\frac{-7\mu_\pi +6\mu_K +\mu_\eta}{3f^2}}\right)
+\left({\frac{3m_\eta^2 +m_\pi^2}{64\pi^2f^2}}\right)
\Biggl( 1+\nonumber \\
&&\qquad\qquad \left[ {\frac{m_\pi^2}{m_K^2-m_\pi^2}}\right]\log\left(
{\frac{m_\pi^2}{m_K^2}}\right)\Biggr) 
+{\frac{(m_\eta^2-m_\pi^2)}{64\pi^2f^2}}\left( 1+\log (m_K^2/\mu^2)\right) 
\nonumber\\
&&\qquad\qquad - {\frac{16B_0}{3f^2}}(m_s-\hat{m})L_5^r
-{\frac{96B_0}{f^2}}(m_s-\hat{m})L_7^r -{\frac{32B_0}{f^2}}
(m_s-\hat{m})L_8^r
\Biggr] \label{mI1eta}\\
&&<0|P_u+P_d|\eta >= { \frac{m^2_\pi f_\pi}{\sqrt{3}\hat{m}}}
\Biggl[ 1+\left( {\frac{-2\mu_\pi +2\mu_K}{f^2}}\right)
-{\frac{16B_0}{3f^2}}(m_s-\hat{m})L_5^r -{\frac{64B_0}{f^2}}(m_s-\hat{m})L_7^r
\Biggr] \label{mlI0eta}\\
&&<0|P_s|\eta >= -{ \frac{m^2_\pi f_\pi}{\sqrt{3}\hat{m}}}
\Biggl[  1+\left( {\frac{\mu_\pi -\mu_\eta }{f^2}}\right)
 - {\frac{16B_0}{3f^2}}(m_s-\hat{m})L_5^r
+{\frac{32B_0}{f^2}}(m_s-\hat{m})L_7^r \nonumber \\
&&\qquad\qquad +{\frac{32B_0}{f^2}}
(m_s-\hat{m})L_8^r 
\Biggr]\ . \label{mseta}
\end{eqnarray}
The cancellation of divergences, and the overall scale-independence of
the expressions for the matrix elements above, provide non-trivial
checks on the calculations, as do the relations
\begin{equation}
<0|\partial_\mu A_f^\mu |P>=<0|2m_fP_f |P>\ +\ {\rm anomalous\ terms}
\label{ward}
\end{equation}
which follow from the Ward identities between the divergence of the
axial currents and the pseudoscalar currents.  (Note that the isospin-breaking 
vacuum-to-$\pi^0,\eta$ matrix elements 
of the axial currents $A_8^\mu$ and $A_3^\mu$
have been evaluated previously by Gasser and Leutwyler\onlinecite{gl85},
and that the anomalous terms in Eq.~(\ref{ward}) cancel when one considers
the combinations which enter $A_3^\mu$ and $A_8^\mu$.  The 
vacuum-to-pseudoscalar matrix elements of the anomalous terms are given
explicitly in Ref.~\onlinecite{DW92}.)
In bringing the expressions for the matrix elements into the forms
displayed above, we have made use of the 1-loop expressions for
$m_\pi^2$ and $f_\pi$\onlinecite{gl85}.  The appearance of the common 
multiplicative factor $m_\pi^2f_\pi /\hat{m}$ in all the matrix 
elements has no physical significance; the equations are simply
written in this form in order to simplify the later calculation of
the mixing parameters relevant to the ``pure'' $\pi^0$ interpolating
field.

\section{Interpolating Fields for the Neutral Pion}

As explained above, the MH $\pi^0$ interpolating field choice
Eq.~(\ref{mhpi0}) is not unique.  In fact, if we consider
\begin{equation}
P(\theta )=(P_u-P_d)+\epsilon (\theta )\left( \cos\theta\, P_8
+\sin\theta\, P_0\right)\label{ptheta}
\end{equation}
where $P_0=(P_u+P_d+P_s)/\sqrt{3}$, the results of Eqs.~(\ref{mlI0pi}),
(\ref{mspi}) show that, in general, 
\begin{equation}
<0|\cos\theta\, P_8+\sin\theta\, P_0|\eta >\not= 0\ .
\end{equation}
As a result, one can, using the construction above, chose $\epsilon (
\theta )$ in Eq.~(\ref{ptheta}) in such a way that 
$<0|P(\theta )|\eta >=0$ for any value of $\theta$.  Any such
$P(\theta )$ can then be employed in an analogue of the MH analysis of 
the three point correlator, in the sense that, in all such analyses,
the unwanted $\eta$ contributions will have been removed by the
corresponding choice of $\epsilon (\theta )$.

The existence of the family of $\pi^0$ interpolating fields above is
potentially useful as a means of investigating features of the 
three-point-function method that 
have been the subject of criticism in the
literature.  The most serious such criticism of the method as a way of
estimating the $\pi NN$ couplings is the objection that it requires
making the assumption of pion dominance in the pseudoscalar channel
all the way out to values of $q^2$ of order $-1\ {\rm GeV}^2$ for
which one can begin to hope that the operator product expansion (OPE)
analysis of the correlator might be valid\onlinecite{ref7}.  The
danger, of course, is that, so far from the pion pole, the contributions
of other resonances ($\eta^\prime$, $\eta (1295)$, $\pi (1300)$, etc.)
might no longer be negligible.  Although MH have not attempted to
investigate this aspect of the method in detail, they argue that their
choice of $\pi^0$ interpolating field (corresponding to $\theta =0$,
and the leading chiral order estimate for $\epsilon (\theta )$ above)
represents the unique one that, at least to leading order, also has
no overlap with the $\eta^\prime$.  At first sight this seems a plausible
argument since, in the $SU(3)_F$ and isospin symmetry limits, 
$P_u-P_d$ couples only to $I=1$ states and $P_8$ only to flavor
octet states.  On closer inspection, however, one can see that
$<0|P(0)|\eta^\prime >\not= 0$\onlinecite{krmnew}, so that there
remain potentially dangerous higher resonance contaminations.  Although
this point is part of a more detailed analysis of higher resonance
contaminations in the three-point-function approach, the results of
which will be presented elsewhere, we will briefly outline the
key aspects of the argument below.  The main point of relevance to
the present discussion is that, although one can, in principle, find
a value for $\theta$ for which at least the vacuum-to-$\eta^\prime$
matrix element vanishes, the information required to actually
determine this value is not available.  Thus, for any particular
choice of $\theta$, one has a coupling of $P(\theta )$ to the
$\eta^\prime$ of unknown strength (in addition to potential
couplings to other higher pseudoscalar resonances), which presents
a serious problem for the three-point-function method.  

One way of dealing with this problem is to investigate the extracted values
of the isospin-breaking $\pi NN$ couplings as a function of $\theta$.
If one finds a region of $\theta$ in which the results are not 
sensitive to variations of $\theta$, then one will have some {\it
a posteriori} justification for assuming that the channel of
interest is, indeed, dominated by the lowest-lying ($\pi^0$) state.
Note that the coupling of $P(\theta )$ to the physical $\pi^0$,
from Eqs.~(\ref{mI1pi})--(\ref{mspi}) above, is, to ${\cal O}(m_d-m_u)$,
$m_\pi^2 f_\pi^2/\hat{m}$, independent of $\theta$.  The 
vacuum-to-$\pi^0,\eta$ matrix elements above provide the essential
input to such an analysis, and we will recast them in a form suitable
for use in such an analysis below.

Before doing so, however, let us briefly outline why the MH field choice,
$\theta = 0$, does not actually remove the coupling to the $\eta^\prime$.
Recall that the quark mass matrix responsible for flavor and isospin
breaking can be decomposed as
\begin{equation}
M=-{\frac{1}{2}}(m_d-m_u)\lambda_3 -{\frac{1}{\sqrt{3}}}(m_s-\hat{m})
\lambda_8
+{\frac{1}{3}}(m_s+2\hat{m})\ .
\label{mdecomp}\end{equation}
To leading order in the isospin-breaking and flavor-breaking
mass differences, therefore, the vacuum-to-$\eta^\prime$ matrix elements
of $P_u-P_d$ and $P_8$ are related simply by the ratio of mass-dependent
coefficients of $\lambda_3$ and $\lambda_8$ in Eq.~(\ref{mdecomp}).
It then follows straightforwardly\cite{krmnew} that, to leading order
in the masses,
\begin{equation}
<0|P(0)|\eta^\prime >=3\theta_0<0|P_8|\eta^\prime >
\end{equation}
where $<0|P_8|\eta^\prime >={\cal O}(m_s-\hat{m})$.  The RHS is thus
non-zero, and ${\cal O}(m_d-m_u)$.  If one considers $\theta\not= 0$,
of course, one has, in addition to the $8_F\times 8_F\rightarrow 1_F$
reduced matrix element which governs $<0|P_u-P_d|\eta^\prime >$ and
$<0|P_8|\eta^\prime >$, the $1_F\times 1_F\rightarrow 1_F$ reduced
matrix element relevant to $<0|P_0|\eta^\prime >$.  One can then
certainly, in principle, find a $\theta$ such that
$<0|P(\theta )|\eta^\prime >=0$.  The problem is that, even to
do so at leading order in the quark masses, one would need to know
the ratio of the two reduced matrix elements above, and this
information is not available.  Moreover, even if it were, this
would not necessarily ensure that, for such a value of $\theta$, the
couplings of the higher resonances were small for the same value
of $\theta$.  We will return to these issues, and to a discussion of the
reliability of the three-point-function method, in more detail in
a future publication\cite{krmnew}.

Let us return, then, to our 1-loop evaluation of $\epsilon (\theta )$.
From Eqs.~(\ref{epspi}), (\ref{mI1eta}), (\ref{mlI0eta}) and (\ref{mseta})
it follows that
\begin{equation}
\epsilon (\theta )=\left[ {\frac{\theta_0}{\cos\theta}}\right]{\frac{N}{D}}
\label{epstheta}
\end{equation}
where
\begin{eqnarray}
N&=&\Biggl[ 1+\left( {\frac{-7\mu_\pi +6\mu_K +\mu_\eta}{3f^2}}\right)
+\left({\frac{3m_\eta^2 +m_\pi^2}{64\pi^2f^2}}\right)
\left( 1+\left[ {\frac{m_\pi^2}{\bar{m}_K^2-m_\pi^2}}\right]\log\left(
{\frac{m_\pi^2}{m_K^2}}\right)\right) 
\nonumber \\
&&\quad
+{\frac{(m_\eta^2-m_\pi^2)}{64\pi^2f^2}}\left( 1+\log (m_K^2/\mu^2)\right) 
- {\frac{16B_0}{3f^2}}(m_s-\hat{m})L_5^r\nonumber \\
&&\quad -{\frac{96B_0}{f^2}}(m_s-\hat{m})L_7^r -{\frac{32B_0}{f^2}}
(m_s-\hat{m})L_8^r
\Biggr] \label{Ntheta} \\
D&=&\Biggl[ 1+\left( {\frac{-3\tan\theta\, \mu_\pi + 2(1+\tan\theta )\mu_K
+(\tan\theta -2)\mu_\eta}{3f^2}}\right) -{\frac{16B_0(m_s-\hat{m})}{3f^2}}L_5^r
\nonumber \\
&&\quad -{\frac{32B_0(m_s-\hat{m})}{f^2}}\tan\theta L_7^r
-{\frac{32B_0(m_s-\hat{m})}{3f^2}}(\tan\theta -2)L_8^r\Biggr]\ .
\label{Dtheta}
\end{eqnarray}
The factor $\theta_0/\cos\theta$ represents the leading order contribution 
to $\epsilon (\theta )$ in the chiral expansion and, of course,
reproduces $\epsilon^{MH}$ (corresponding to $\theta = 0$) at this
order.  
The remaining factors represent the next-to-leading order
corrections.  We will discuss their magnitude below.
One should note that,
as $\theta\rightarrow \pi /2,\ 3\pi /2$, the denominator of
Eq.~(\ref{epstheta}) tends to a finite value involving the chiral log
and fourth order LEC terms.  This means that, near such values of
$\theta$, $\epsilon (\theta )$ becomes large, and with significant
uncertainties (associated with uncalculated next-to-next-to-leading
order contributions to the denominator).  As such, the corresponding
interpolating field choices, though in principle 
reasonable, are not useful practically.  In order to avoid these
regions of $\theta$ it is convenient to consider a re-parametrization
of the family of interpolating fields in which one writes
\begin{eqnarray}
P(\alpha )&=&(P_u-P_d)+\epsilon (\alpha )\Bigl[ \alpha (P_u+P_d)+(1+\alpha )P_s
\Bigr]\nonumber\\
&=& (P_u-P_d)+\epsilon (\alpha ){\frac{1}{\sqrt 3}}\Bigl[ -P_8+(3\alpha +1)
P_0\Bigr]\ ,
\label{alphafamily}
\end{eqnarray}
where, as usual, $\epsilon (\alpha )$ is to be chosen for each $\alpha$
in such a way that 
\begin{equation}
< 0|P(\alpha )|\eta >=0\ .
\label{constraint}
\end{equation}
The second line of Eq.~(\ref{alphafamily}) shows that
such a form for $P(\alpha )$
represents a simple reparametrization of $P(\theta )$, in which a factor
of $-{\sqrt 3}\cos \theta$ has been absorbed into $\epsilon$ and 
$\tan\theta$ has been rewritten as $3\alpha +1$.  The regions of
$\theta$ we wish to avoid then correspond to $\alpha\rightarrow\infty$.
One then easily shows that Eq.~(\ref{constraint}) implies
\begin{eqnarray}
\epsilon(\alpha )&=& -{\sqrt 3}\theta_0\Biggl[ 1
+\left(
{\frac{-(10+9\alpha )\mu_\pi +6(1+\alpha )\mu_K +(4+3\alpha )\mu_\eta}{3f^2}}
\right)\nonumber \\
&&\qquad -{\frac{32}{3f^2}}
(4+3\alpha )\left( \bar{m}_K^2-m_\pi^2 \right)\left( 3L_7^r+L_8^r\right)
+\left({\frac{3m_\eta^2 +m_\pi^2}{64\pi^2f^2}}\right)
\Biggl( 1+\nonumber \\
&&\qquad \left[ {\frac{m_\pi^2}{\bar{m}_K^2-m_\pi^2}}\right]\log\left(
{\frac{m_\pi^2}{\bar{m}_K^2}}\right)\Biggr)
+{\frac{(m_\eta^2-m_\pi^2)}{64\pi^2f^2}}\left( 1+\log (m_K^2/\mu^2)\right) 
\Biggr]\ \nonumber \\
&=&  -{\sqrt 3}\theta_0\, \left[ 1.299 + 0.0894\, \alpha\right]\ ,
\label{epsalpha} 
\end{eqnarray}
where, in arriving at Eq.~(\ref{epsalpha}), we have made an expansion
of the denominator in Eq.~(7) so that the whole expression in now explicitly
consistent to 1-loop order.  The reparametrization has been chosen such
that, at leading order in the chiral expansion, $\epsilon (\alpha )$
reduces to $-{\sqrt 3}\, \theta_0$, so the leading $1$ in the
square brackets in Eq.~(\ref{epsalpha}) represents the tree-level
contribution.  The remaining terms are the 1-loop corrections and,
as expected, are not small.
This can be seen from the last line of Eq.~(\ref{epsalpha}), where
the deviation of the term $1.299+0.894\, \alpha$ from $1$ given
the size of the 1-loop corrections.
For reference sake we note that the MH choice
$\theta =0$ corresponds to $\alpha =-1/3$, and the choice in which
there is no $P_s$ content to the interpolating field corresponds to
$\alpha =-1$.  (To completely convert back to $\theta$ 
notation for the MH case, one
must also absorb the constant $-1/{\sqrt 3}$ back into the definition
of $\epsilon$.)
The enhancements due to next-to-leading order corrections are, in
these cases, $1.27$ and $1.21$, respectively.  Amusingly, the choice
$\alpha =-4/3$ removes the LEC contributions completely from the
expression for $\epsilon$.

\section{Summary}

We have evaluated the leading ${\cal O}(m_d-m_u)$ strong isospin-breaking
contributions to the matrix elements $<0|P_f|\pi^0,\eta >$ at
next-to-leading order in the chiral expansion.  The results are given
in Eqs.~(\ref{mI1pi})--(\ref{mseta}).  We have also showed how to
use this information to construct a family of $\pi^0$ interpolating fields
which could, in future, be employed to investigate the reliability of
analyses which attempt to extract the isospin-breaking $\pi NN$
couplings via the three point function method in QCD sum rules.  The
isospin-breaking parameters, $\epsilon (\alpha )$, required in this
construction in order to remove unwanted $\eta$ contaminations
from the extracted $\pi^0$ couplings, were shown to receive 
significant contributions at next-to-leading order.  As a result,
if one wishes to remove such contamination, it is essential that the
full next-to-leading order expressions, given in Eq.~(\ref{epsalpha}),
be used.

\acknowledgements

The authors acknowledge the continuing financial support of the Natural 
Sciences and Engineering Research Council of Canada.  KM would also like
to acknowledge useful conversations with Terry Goldman, Derek Leinweber,
Tony Williams and Mike Birse, as well as the hospitality of the
Special Research Center for the Subatomic Structure of Matter at the
University of Adelaide, where much of this work was performed.

\begin{figure}[htb]
  \centering{\
    \epsfig{angle=0,figure=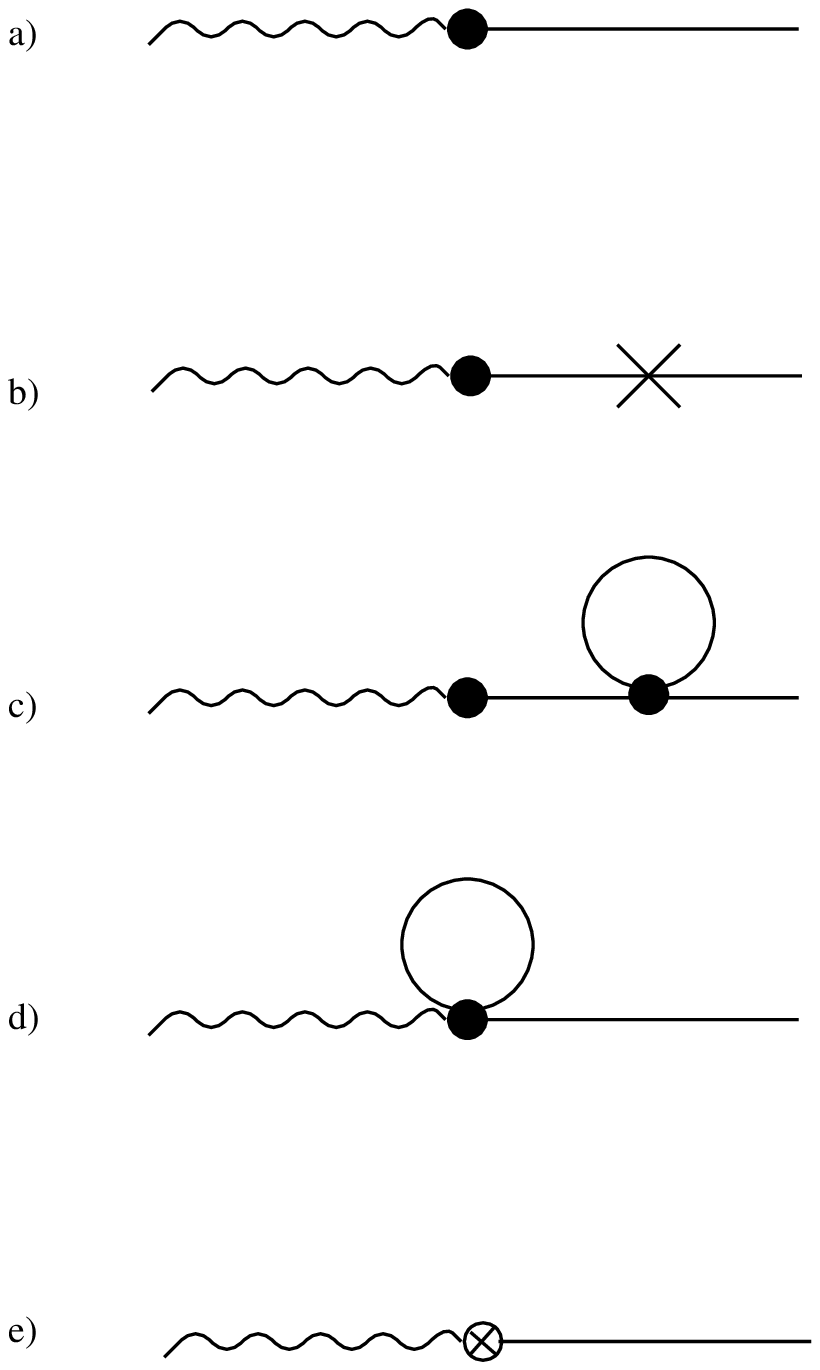,height=12.cm}
      \vskip .5in         }
\parbox{130mm}{\caption
{Graphs contributing to the vacuum-to-$\pi^0$, $\eta$ matrix elements.
The wavy lines represent the pseudoscalar currents,
the solid dots the vertices generated by the ${\cal O}(q^2)$ part
${\cal L}_{eff}$, the `X' and the circle enclosing an `x'
vertices generated by the ${\cal O}(q^4)$ part of ${\cal L}_{eff}$.}
% }
\label{figone} }
\end{figure}


\begin{references}
\bibitem{svz}M.A. Shifman, A.I. Vainshtein and V.I. Zakharov, Nucl.
Phys. {\bf B147}, 385, 448 (1979).
\bibitem{ref1}L.J. Reinders, H. Rubinstein and S. Yazaki, Nucl.
Phys. {\bf B213}, 109 (1983).
\bibitem{ref2}L.J. Reinders, Act. Phys. Pol. {\bf 15}, 329 (1984).
\bibitem{ref3}L.J. Reinders, H. Rubinstein and S. Yazaki, Phys. Rep.
{\bf 127}, 1 (1985).
\bibitem{ref4}S. Narison and N. Paver, Phys. Lett. {\bf B135}, 159 (1984).
\bibitem{ref5}S. Narison, {\it QCD Spectral Sum Rules} (World Scientific,
Singapore, 1989).
\bibitem{ref6}H. Shiomi and T. Hatsuda, Nucl. Phys. {\bf A594}, 294 (1995).
\bibitem{ref7}M. Birse and B. Krippa, Phys. Lett. {\bf B373}, 9 (1996).
\bibitem{mnsrev}G.A. Miller, B.M.K. Nefkens and I. Slaus, Phys. Rep.
{\bf 194}, 1 (1990).
\bibitem{ght}T.~Goldman, J.~A.~Henderson and A.~W.~Thomas,
Few-Body Systems {\bf 12},123 (1992).
\bibitem{pw93}J.~Piekarewicz and A.~G.~Williams, Phys.\ Rev.\ {\bf C47}, 
R2461 (1993); J.~Piekarewicz, Phys.\ Rev.\ {\bf C48}, 1555 (1993), 
Phys. Lett. {\bf B358}, 27 (1995).
\bibitem{gtw93}G.~Krein, A.~W.~Thomas, and A.~G.~Williams, Phys.\ Lett.\
{\bf B317}, 293 (1993); H. O'Connell, B. Pearce, A.W. Thomas and A.G.
Williams, Phys. Lett {\bf B336}, 1 (1994);
H. O'Connell, B. Pearce, A.W. Thomas and A.G.
Williams, Phys. Lett. {\bf B354}, 14 (1995);
H. O'Connell, A.G. Williams, M. Bracco and
G. Krein, Phys. Lett. {\bf B370}, 12 (1996).
\bibitem{mg94}K.~Maltman and T.~Goldman, Nucl.\ Phys.\ {\bf A572}, 682 (1994).
\bibitem{chm}C.-T. Chan, E.M. Henley and T. Meissner, Phys. Lett. {\bf B343},
7 (1995).
\bibitem{hhkm}T. Hatsuda, E.M. Henley, T. Meissner and G. Krein,
Phys. Rev. {\bf C49}, 452 (1994).
\bibitem{krmps}K.~Maltman, Phys.\ Lett.\ {\bf B313}, 203 (1993); Phys.
Lett. {\bf B351}, 56 (1995).
\bibitem{krm12}K. Maltman, Phys. Rev. {\bf D53}, 2563, 2573 (1996); Phys.
Lett. {\bf B362}, 11 (1995).
\bibitem{kent}K.~L.~Mitchell, P.~C.~Tandy, C.~D.~Roberts and
R.~T.~Cahill, Phys.\ Lett.\ {\bf B335} (1994) 282; M.R. Frank and P.C. Tandy,
Phys. Rev. {\bf C49}, 478 (1994); M.R. Frank, Phys. Rev. {\bf C51}, 987
(1995); K.L. Mitchell and P.C. Tandy, Phys. Rev. {\bf C55}, 1477 (1997).
\bibitem{chptcsb}R. Urech, Phys. Lett. {\bf B355}, 308 (1995); H. Neufeld
and H. Rupertsberger, Z. Phys. {\bf C68}, 91 (1995).
\bibitem{ghp}S. Gardner, C.J. Horowitz and J. Piekarewicz, Phys. Rev.
Lett. {\bf 75}, 2462 (1995); Phys. Rev. {\bf C53}, 1143 (1996).
\bibitem{vkgf}U. van Kolck, J. Friar and T. Goldman, Phys. Lett. {\bf B371},
169 (1996).
\bibitem{ijl}M.J. Iqbal, X.-M. Jin and D.B. Leinweber, Phys. Lett. {\bf B367},
45 (1996); Phys. Lett. {\bf B386}, 55 (1996).
\bibitem{cm95}T.D. Cohen and G.A. Miller, Phys. Rev. {\bf C52}, 3428 (1995).
\bibitem{mow}K. Maltman, H. O'Connell and A.G. Williams, Phys. Lett. {\bf
B376}, 19 (1996).
\bibitem{hm95}T. Meissner and E.M. Henley, nucl-th/9601030,
Phys. Rev. {\bf C}, in press.
\bibitem{gl85}J.~Gasser and H.~Leutwyler, Nucl.\ Phys.\ {\bf B250} (1985) 465
\bibitem{weinberg79}S. Weinberg, Physica {\bf 96A}, 327 (1979).
\bibitem{bkm95}V. Bernard, N. Kaiser and U.-G. Meissner, Int. J. Mod.
Phys. {\bf E4}, 193 (1995).
\bibitem{DW92}J.F. Donoghue and D. Wyler, Phys. Rev. {\bf D45}, 892 (1992).
\bibitem{krmnew}K. Maltman, ``Higher Resonance Contamination of
$\pi NN$ Couplings Obtained Via the Three-Point Function Method in
QCD Sum Rules'', York University preprint, June 1997.
\end{references}
\end{document}